\documentstyle[twocolumn,prl,aps,graphicx,amsmath,amssymb,inputenc,floats]{revtex}

\inputencoding{ansinew}

\newcommand{\WN}{\mbox{cm$^{-1}$}}
\newcommand{\degree}{^\circ}
\newcommand{\MB}{MgB$_{2}$}
\newcommand{\etal}{\textit{et al.}}
\newcommand{\Tc}{T_c}

\narrowtext

\begin{document}

\draft


\date{received 21 May 2001}

 \twocolumn[\hsize\textwidth\columnwidth\hsize\csname
 @twocolumnfalse\endcsname

\title{MgB$_2$ under pressure: phonon calculations, Raman spectroscopy,\\ and
optical reflectance}

\author{K. Kunc\cite{byline1}, I. Loa\cite{byline2},
K.~Syassen, R. K. Kremer, and K. Ahn}

\address{Max-Planck-Institut f\"ur Festk\"orperforschung,
Heisenbergstrasse 1, D-70569 Stuttgart, Germany}

\maketitle

\begin{abstract}
The effect of pressure on optical phonon frequencies of MgB$_2$ has been
calculated using the frozen-phonon approach based on a pseudopotential method.
Gr\"{u}neisen parameters of the harmonic mode frequencies are reported for the
high-frequency zone-center $E_{2g}$ and $B_{1g}$ and the zone-boundary $E_{2u}$
and $B_{2u}$ modes at $A$. Anharmonic effects of phonon frequencies and the
implications of the calculated phonon frequency shifts for the pressure
dependence of the superconducting transition temperature of MgB$_2$ are
discussed. Also reported are Raman and optical reflectance spectra of MgB$_2$
measured at high pressures. The experimental observations in combination with
calculated results indicate that broad spectral features we observed in the Raman
spectra at frequencies between 500 and 900 cm$^{-1}$ cannot be attributed to
first-order scattering by zone-center modes, but originate in part from a
chemical species other than MgB$_2$ at the sample surface and in part from a
maximum in the MgB$_2$ phonon density of states. Low-temperature Raman spectra
taken at ambient pressure showed increased scattering intensity in the region
below 300
\WN.
\end{abstract}


] \narrowtext

\section{Introduction}

Magnesium diboride, MgB$_2$, was recently discovered to exhibit a superconducting
transition at 39 K \cite{nagamatsu}, which is by far the highest $T_c$ for a
binary compound. The light atomic masses in MgB$_2$ enhance the phonon
frequencies which set the scale for $T_c$ in BCS theory. A variety of
experimental observations (e.g., isotope effects~\cite{budko} and scanning
tunneling measurements of the superconducting gap
\cite{RSV01pv2,KIKC01p,SFM01pv2}), indeed indicate phonon-mediated (BCS)
superconductivity, and theory finds medium or strong electron-phonon coupling
\cite{KMBA01,AP01p,KDJA01pv3,YGLB01p}. The coupling is predicted to be
particularly strong for the zone-center optical phonon of $E_{2g}$ symmetry and
the related phonon branch connecting to the zone-boundary $E_{2u}$ mode at the
$A$ point (direction along the $c$ axis of the AlB$_2$-type structure)
\cite{AP01p,KDJA01pv3,YGLB01p}.

The effect of pressure on the superconducting properties of MgB$_2$ has been
studied by several
groups.~\cite{STII01,MNRR01,LMC01pv2,THSH01pv2,LMC01ap,AGK01unpub} While there is
some scatter in the experimental pressure coefficients all studies show a
decrease of $T_c$, at  rates between 0.7 and 2.0 K/GPa. Lattice parameters under
pressure have been determined by neutron and x-ray diffraction
\cite{VSHY01p,prassides,JHS01p,GSGH01pv2}. An {\it ab initio} calculation of the
pressure-volume relation and the optimized $c/a$ axial ratio under pressure
\cite{LS01} is in good agreement with the experimental data. It was concluded
that the decrease in $T_c$ is not driven by a change in the electronic density of
states near $E_F$ but by an increase in phonon frequencies, specifically of the
strongly coupled $E_{2g}$ mode \cite{LS01}.

The frequency of the $E_{2g}(\Gamma)$ phonon is expected between 470 and 665
cm$^{-1}$, according to recent calculations
\cite{KMBA01,GGFA01p,KDJA01pv3,BHR01p,YGLB01p}. So far, this phonon mode has not
been identified unambiguously by neutron inelastic scattering
\cite{YGLB01p,OGKH01p,sato} or in Raman experiments
\cite{GSGH01pv2,BHR01p,CKIL01p,MMRL01pv2,HGPP01p}. A Raman spectrum by Bohnen
\etal \cite{BHR01p} indicates a broad asymmetric feature near 580~\WN, while
other authors \cite{GSGH01pv2,MMRL01pv2,HGPP01p} report an even broader feature
(260--300~\WN\ width) centered near 620~\WN. Contradictorily, compliance with the
selection rules for a $E_{2g}$ mode has been reported \cite{HGPP01p} as well as
complete breakdown of the same, attributed to resonance effects \cite{MMRL01pv2}.
The 620-\WN\ feature was observed to increase in frequency under pressure, with
an unusually large Gr\"{u}neisen parameter \cite{GSGH01pv2}. It was generally
assigned to first-order zone-center phonon scattering, highly broadened by strong
electron-phonon coupling or anharmonic effects, but other interpretations were
also considered.

In this paper we present a theoretical investigation of the pressure dependence
of phonon frequencies of MgB$_2$. As no phonon has been, so far, unambiguously
identified experimentally, the theoretical methods gain in importance -- in
particular the {\it ab initio} approaches, such as the frozen phonon method which
is employed in this work. Compared to linear response theory, the frozen phonon
method has the disadvantage that it is not very efficient in generating an
overall picture of the phonon dispersions throughout the Brillouin zone, but is
restricted to calculation of a few high-symmetry modes. One of the advantages of
the frozen phonon method, however, is that it yields a clear picture of
anharmonic effects, which in earlier calculations \cite{YGLB01p} were found to be
particularly large for the atomic displacements corresponding to the $E_{2g}$
mode of MgB$_2$. The immediate interest in the shifts of phonon frequencies is in
the context of interpreting the pressure dependence of $T_c$. Furthermore, the
calculated pressure dependences of phonon frequencies can in principle be useful
for the identification of spectral features in Raman spectra.

We also report here Raman and optical reflectance measurements of
MgB$_2$ as a function of pressure at room temperature. Based on the
pressure dependence of Raman spectra and the optical reflectance and
by taking into account the results of the phonon frequency
calculations we argue that the observed Raman features
are not due to zone-center first-order Raman scattering from
MgB$_2$. The interpretation we suggest for our Raman results considers a
superposition of a density-of-states peak of MgB$_2$ and scattering by
a different chemical species present at the surface of as-grown
samples. We also take a look at zero-pressure Raman spectra measured
as a function of temperature in order to check whether it is possible
to observe a pair-breaking excitation below $\Tc$.

The paper is organized as follows: the symmetry analysis of all phonon modes at
the $\Gamma$ and $A$ point is presented in Section \ref{sec:symm}. Details of the
theoretical method and the calculated results are presented and discussed in
Section \ref{sec:calc}. Experimental details and results are given in Section
\ref{sec:exp}, followed by conclusions in Section \ref{sec:concl}.

\section{Symmetry analysis}
\label{sec:symm}

The frozen phonon calculations described below require knowledge of
the phonon displacement patterns. We thus start with the symmetry
analysis of the vibrational spectra. The point group of the AlB$_2$
structure is $D_{6h}$. A factor-group analysis yields the
decomposition of the coordinate representation

\begin{eqnarray}
    B_{1g} + E_{2g} + 2A_{2u} + 2E_{1u} \ \ {\rm at}\ \  \Gamma
\end{eqnarray}
    and
\begin{eqnarray}
    A_{1g} + E_{1g} + A_{2u} + B_{2u} + E_{1u} + E_{2u} \ \ {\rm at}\ \  A
\end{eqnarray}

\noindent with, at the point $\Gamma$, one $A_{2u}$ and one
$E_{1u}$ referring to the rigid translations ($\omega=0$). Constructing then
projection operators we find the bases spanning the irreducible representations
in Eqs. 1 and 2 -- i.e., the respective phonon displacement patterns at $\Gamma$
and $A$, respectively. They are shown in Figs. \ref{fig1} and \ref{fig2}, with
the modes arranged (columnwise) in the order of decreasing MgB$_2$
eigenfrequencies (which are obtained later, in Section II.C). At high pressures
this ordering may, of course, be altered. The patterns in Fig. \ref{fig2} are
arranged so as to be the ``end-points'' of those $\Gamma$--$A$ branches of the
phonon dispersion that start with the corresponding pattern in Fig. \ref{fig1}.
This means that e.g.\ the three patterns in the last column of Fig. \ref{fig2}
are the end-points of the acoustic branches.

The patterns at $A$ alternate the $+u$ and $-u$ displacements from one layer to
another. In addition, for the two highest frequency modes we notice that (to
within the $\pm$ alternation) the patterns of a layer are the same as at
$\Gamma$.

\section{Calculations}
\label{sec:calc}

\subsection{Theoretical Method}

The frozen phonon approach starts from the {\it ab initio} evaluation of the
total energy $E^{tot}$ of the solid with frozen-in atomic displacements. The
energy is evaluated using the density functional theory (DFT) within the
generalized gradient approximation (GGA) \cite{gga}. We work in a plane-wave
basis and use pseudopotentials. For the actual calculations we employed the
VASP codes \cite{vasp1,vasp2,vasp3,vasp4}, and the ultra-soft
Vanderbilt-type pseudopotentials \cite{vasp5} were supplied by Kresse and
Hafner \cite{vasp6}. The pseudopotential we chose for Mg treats the
semi-core states explicitely, i.e., Mg is represented with 8 valence
electrons $2p^63s^2$, and the non-linear core correction (NLCC) \cite{nlcc}
is applied as well. The calculations are carried out with the plane-wave
cutoff energy of 33.6 Ry, and the Brillouin zone sampling is based on
$\Gamma$-centered uniform meshes $16 \times 16 \times 14$ or  $16
\times 16 \times 8$ -- the latter one applied in the calculations with
the doubled unit cell for frozen phonons at the Brillouin zone boundary.
As the system is metallic, the $\vec k$-space integration with the
incompletely filled orbitals uses the tetrahedron method \cite{tetra}
with Bl\" ochl's corrections \cite{bloechl}. The above meshes divide
the Brillouin zone into 21504 or 12288 tetrahedra; the number of {\it
irreducible} tetrahedra is lower by a factor 2 to 6, depending on the
symmetry of the frozen-in phonon.

\subsection{Structural Properties}

For eleven volumes between 25.0 \AA$^3$ and 29.6 \AA$^3$, we calculated
$E^{tot}(V)$ with different $c/a$ values while keeping the volume $V$ constant
and, for every chosen $V$, found the optimized structure, i.e. the corresponding
$a(V)$, $c(V)$, $c/a(V)$, and $E(V)$. The latter data were fitted by the
Murnaghan relation for $E(V)$\cite{Murnaghan44} which provided the static
equilibrium\cite{foot1} as well as the bulk modulus $B_0$ and its pressure
derivative $B'$. The results are summarized in Table \ref{tab1}. The calculated
equilibrium volume ($V_0$ = 28.663 \AA$^3$) compares well with the experimental
value $V_0$ = 28.917(1) \AA$^3$ (Ref.~\cite{JHS01p}, at 37 K) meaning that we are
-0.9\% off in $V_0$. The individual lattice parameters are -0.6\% off the
experiment for $a_0$, +0.2\% for $c_0$, and +0.8 \% for $c_0/a_0$. The variation
of $c/a$ under pressure has been translated into a quadratic function of
$(1-V/V_0)$ and its coefficients are given in Table \ref{tab1} as well. The
present results are also in good agreement with those of a previous calculation
\cite{LS01} (see Table~\ref{tab1}).

A limited testing of different pseudopotentials revealed that treating the
magnesium 2$p^6$ electrons as valence states is {\it not} essential for getting
the correct equilibrium, but using the GGA (rather than the LDA) {\it is}
\cite{foot2}. With the value of $c/a = 1.14$ the MgB$_2$ is well inside the
stability region of the AlB$_2$ structures \cite{pearson}. A rather high pressure
would probably be required to induce a phase transition, possibly towards the
UHg$_2$-type structure (isostructural to AlB$_2$, but $c/a$ less than $\sim$ 0.8)
or a variant thereof, as is observed in other AlB$_2$-type intermetallic
compounds \cite{Schwarz}.

The subsequent calculations of phonon frequencies are performed at the
{\it calculated} equilibrium, and at volumes related to pressure by
the Murnaghan relation

\begin{equation}\label{murn}
P(V)~=~{B_0 \over B^{'}} \left [ \left ( {V/V_{0}} \right ) ^{-B^{'}}
- 1 \right ]\: , \end{equation}

\noindent which uses the {\it calculated} $V_0, B_0, B'$
quoted in Table \ref{tab1}.

\subsection{Phonon frequencies}

In this work we are mainly concerned with the two highest frequency
modes $B_{1g}(\Gamma)$ and $E_{2g}(\Gamma)$ and their $A$-point
counterparts $B_{2u}(A)$ and $E_{2u}(A)$. For comparison with previous
works we calculated the $A_{2u}(\Gamma)$ and $E_{1u}(\Gamma)$ as well.
The branch $E_{2g}(\Gamma)-E_{2u}(A)$ in the phonon dispersion is the
one that exhibits the strongest electron-phonon coupling
\cite{KDJA01pv3}, and the $B_{1g}(\Gamma)$ is the $z$-displacement
analogue of the $E_{2g}(\Gamma)$.

We determine the vibrational eigenfrequencies using the frozen phonon
method \cite{frphon}, while paying particular attention to obtaining
the {\em harmonic} part of the ab initio calculated energy differences
that, inevitably, have to be probed by  {\it finite} displacements
$u$. Thus for each pattern, atoms are given 5 different displacements
ranging from $u/a$ = 0.010 to 0.035, and the (six) calculated energy
values $E(u)$ are fitted with a quartic polynomial yielding

\begin{equation} \label{eqeu}
  \Delta E_{tot}(u) =  a_2 (u/a)^2 + a_3 (u/a)^3 + a_4 (u/a)^4
\end{equation}

\noindent As an example we show in Fig. \ref{fig3} the variation of
$E_{tot}(u)$ for the E$_{2g}(\Gamma$) mode \cite{foot3}. The harmonic
part of $\Delta E_{tot}$ is then retained for determination of the
phonon energy:

\begin{equation} \label{eqome}
{\omega^2 \over 2} \sum M_{\kappa} u_{\kappa}^2
                               = E^{harm}(u \ne 0) - E^{harm}(u = 0)
\end{equation}

\noindent For the four phonons listed above the procedure is repeated
at various volumes. Although straightforward, this procedure calls for a
few explanations.

The frozen-in phonon displacements of Fig. \ref{fig1} conserve the
translational symmetry of the elementary unit cell defining the
MgB$_2$ structure. With the zone boundary modes given in Fig.
\ref{fig2} the total energy calculations as a function of displacement
$u$ are to be done on a unit cell obtained by {\it doubling} the
elementary cell (viz. along the [001] direction), which then comprises
two formula units of MgB$_2$. The frozen-in displacements also lower
the rotational symmetry of the system: starting from the $D_{6h}$,
which is the point-group of the MgB$_2$ lattice, we end with,
respectively, $D_{3d}$, $D_{2h}$, $C_{2h}$ when the $B_{1g}(\Gamma)$,
$E_{2g}(\Gamma)$-(a,b) displacements are imposed, and $D_{3h}$,
$D_{2h}$, $D_{2h}$ with the $B_{2u}(A)$, $E_{2u}(A)$-(a,b)
displacements, respectively.

As the displacements $u$ are finite, a larger or smaller degree of anharmonicity
will always be present, but some restrictions on the latter can be read out from
symmetry. With all patterns at the zone boundary, replacing $+u$ by $-u$ has the
same effect as shifting the phonon distorted structure by the lattice constant
$c$, and it leaves the $E(u)$ invariant. Consequently, the cubic (and, generally,
odd-power-) anharmonicity is forbidden in all zone-boundary modes. Inspecting, in
turn, the $\Gamma$-displacement patterns we realize that for nearly all modes the
$+u \leftrightarrow -u$ substitution is equivalent to applying a mirror
reflection or $C_2$ rotation -- an operation which, in spite of the lowered
symmetry, is still present in the point groups of the phonon-modulated structures
-- and which thus leaves invariant the structure and its energy $E(u)$. This
forbids the cubic term in the expansion of $E(u)$ as well. Among the 9 modes in
Fig. \ref{fig1} the $E_{2g}(\Gamma)$-(a) pattern is the only one in which all
anharmonic terms are allowed and in which the cubic anharmonicity is to be
expected.

For evaluation of phonon frequencies from $\Delta E_{tot}(u)$ we used the
atomic masses of B and Mg that correspond to the natural isotope
distribution: 10.8 and 24.3 a.m.u.; this is the usual approach in the spirit
of the virtual crystal approximation. We may keep in mind that for a
$^{10}$B-enriched sample the frequencies of all the modes involving boron
would be higher by 4\%.

Table \ref{tab2} summarizes the results of our frozen phonon
calculations at different volumes:  the variations with volume or
pressure are described by the mode Gr\" uneisen parameter $\gamma_{0}=
- \left[d \ln \omega / d \ln V\right]_{V_0}$ or the coefficients $a$
and $b$ of a quadratic fit (see the caption). Figure \ref{fig4} shows
the calculated phonon frequencies as a function of volume.

We estimate that the calculated harmonic frequencies are uncertain to
$\sim$1--3\% (depending on the mode); this is the ``instrumental precision'' of
the determination of total energies and the uncertainty following from the error
analysis in fitting the variations $E(u)$ by polynomials.

\subsection{Anharmonicity and the $E_{2g}(\Gamma)$ mode}

We note that the five displacements $u/a$ quoted above range from one
half to approximately the double of the r.m.s. displacement
\cite{AP01p}, and one expects vibrations in both harmonic and
anharmonic regimes. The effort of removing from $\Delta E_{tot}$ the
anharmonic contributions yields, in turn, an insight into the
anharmonicities.

The phonons exhibiting the strongest electron-phonon coupling are those of the
$E_{2g}(\Gamma) -  E_{2u}(A)$ branch \cite{KDJA01pv3}. The $E_{2g}(\Gamma)$ has
already received considerable attention
\cite{KMBA01,GGFA01p,KDJA01pv3,BHR01p,YGLB01p}. Being doubly degenerate, it is
materialized by {\it two} distinct displacement patterns which are labeled (a)
and  (b) in Fig.~\ref{fig1}. They both should lead to the same eigenfrequency
(the same coefficient $a_2$ in Eq. \ref{eqeu}) and, in this sense, they are
equivalent -- as long as we stay within the harmonic approximation, i.e., within
the standard group-theory reasonings. Once we step out of the harmonic
approximation, the displacements (a) and  (b) will become two unrelated patterns
which are described by different anharmonic coefficients. This gets confirmed by
numerical calculations  and we obtained $a_2$ = 107 eV/cell with both (a) and (b)
patterns and, somewhat unexpectedly, also $a_4$ = 45800 eV/cell equal in {\it
both} patterns. The two modes are nevertheless distinct in the cubic term, which
is absent in the (b) pattern ($a_3 = 0$) and given by $a_3$ = -1644 eV/cell in
the (a)-pattern.

It is not difficult to understand why the mode $E_{2g}(\Gamma)$-(a)
exhibits a strong cubic anharmonicity: giving the boron atoms a $+u$
or $-u$ displacement, the main contribution to $\Delta E_{tot}$ comes
from stretching or compressing the B--B bonds -- which, obviously, has
very different energetic cost. In the calculations, the cubic
anharmonicity of the mode is dealt with easily: after obtaining the
$\Delta E_{tot}$ with the displacements $+u$ and $-u$ we take an
average, and are left with $\Delta E_{tot}$ carrying quartic
anharmonicity only. Using the pattern (b) instead of (a) is
technically somewhat simpler, because the averaging of the
contributions from the stretched and compressed bonds is ``built in''
within the same displacement pattern.

A striking feature of the $E_{2g}(\Gamma)$ mode (one which already has been
noticed by other authors \cite{YGLB01p}) is its strong anharmonicity $a_4$;
it is larger by an {\it order of magnitude} than that of the
$B_{1g}(\Gamma)$ mode. This is equally true for the corresponding mode
$E_{2u}(A)$ at the Brillouin zone boundary and, apparently, holds for the
whole branch $E_{2g}(\Gamma) - E_{2u}(A)$. The ratio of the quartic and
harmonic energies, at any given displacement $u$, is $a_4/a_2^2$ = 4
eV$^{-1}$ for $E_{2g}(\Gamma)$ and 3.5 for $E_{2u}(A)$. Yildirim
\etal\ \cite{YGLB01p} arrived at an even stronger anharmonicity of the
$E_{2g}(\Gamma)$ mode: $a_4/a_2^2 \approx$ 8 eV$^{-1}$.

Large anharmonicity is consistent with a strong variation of the
phonon frequency with volume and the mode Gr\" uneisen parameters of
$E_{2g}(\Gamma)$ and $E_{2u}(A)$ turn out to be much larger than those
of the other modes (see Table \ref{tab2}) -- and, in any case,
unusually large: 2.5 and 2.8, respectively. The corresponding
frequency shifts $\Delta \omega  \propto \gamma \omega$ are
approximately the same, which means that the whole branch
$E_{2g}(\Gamma)$ -- $E_{2u}(A)$ is shifting ``rigidly'' when
compressed -- unlike the branch $B_{1g}(\Gamma) - B_{2u}(A)$ which
becomes more dispersive under pressure.

It is interesting to note that for the $E_{2g}(\Gamma)$ and $E_{2u}(A)$ modes the
quartic coefficient $a_4$ of Eq. \ref{eqeu} (not given in Table \ref{tab2}) turns
out to be approximately constant and independent of volume, in the entire range
of volumes shown in Fig. \ref{fig4}; this means that, under pressure, the quartic
energy goes as the inverse fourth power of the lattice constant $a$.

A well known  effect of anharmonicity is in broadening the spectral
features in phonon spectroscopies and in shifting them by some amount
$\Delta(\omega)$ with respect to the ideal $\delta(\omega' -
\omega)$-like response of the perfectly harmonic system. The intricate
details of the variation $\Delta(\omega)$ are beyond the scope of this work.
Yildirim \etal\cite{YGLB01p} report that for the E$_{2g}(\Gamma))$ mode the
$\Delta(\omega)$ amounts to a frequency shift by 17 - 25\% above  the harmonic
frequency. The estimate is based on their results for $a_2$, $a_4$,
$\omega(E_{2g}(\Gamma))$= 486 cm$^{-1}$  and employs two different ways of
``renormalizing'' the harmonic frequency. Taking into account that the anharmonic
shift, in the lowest order of perturbation theory, is proportional to the ratio
$a_4/a_2^2$ and to the phonon energy $\hbar
\omega$ itself \cite{kkrz},  a simple rescaling of  the Yildirim \etal\ results
suggests, for our case, a 9 to 13\% upshift of the $E_{2g}(\Gamma)$-frequency;
similarly, the shift in $E_{2u}(A)$ can be estimated at  7 to 10\% (see the
hatched areas in Fig.~\ref{fig4}). Treating in the same way also the calculated
harmonic phonon frequencies in compressed unit cells we estimate that these
shifts reduce the Gr\"uneisen parameters of the two modes from 2.5 and 2.8 to
$\gamma_0 =$ 2.0 -- 2.2 and $\gamma_0 =$ 2.3 -- 2.5, respectively.

\subsection{Comparison with other calculations}

There have been several calculations of zone-center vibrations
\cite{KMBA01,GGFA01p,YGLB01p} and phonon dispersion
\cite{KDJA01pv3,BHR01p,YGLB01p}. The results obtained for modes at $\Gamma$ are
compared in Table \ref{tab3}. The most conspicuous feature of the comparison is
the large spread of the results for the $E_{2g}(\Gamma)$ mode -- the calculated
frequency varies from 470 cm$^{-1}$ \cite{KMBA01} to 665 cm$^{-1}$ \cite{GGFA01p}
-- although otherwise good agreement between different authors is found for the
other $\Gamma$ modes.

Bohnen \etal\cite{BHR01p} attribute the disagreement in the
frequency of the $E_{2g}(\Gamma)$ mode to an insufficiently
converged $\vec k$-point sampling in some of the calculations.
In our case we have ensured that full convergence is reached.

Large anharmonicity of the $E_{2g}(\Gamma)$ was proposed as a possible cause
of the discrepancies by Yildirim \etal\ \cite{YGLB01p}. The proper
``separation'' of the harmonic part in the calculated $\Delta E_{tot}$
indeed matters: evaluating the frozen phonon energy at $u$ = 0.057 \AA \
(=the r.m.s. displacement) we obtain, for the $E_{2g}(\Gamma)$-(a) mode, the
harmonic term in Eq. \ref{eqeu} of 37.84 meV, the cubic 10.57 meV and the
quartic 5.33 meV. Billing the quartic contribution into the harmonic energy
would shift the frequency up by as much as 7\%, assuming that the cubic term
has been correctly eliminated (else the frequency would be shifted by
another $\pm 14$\%).

When linear response theory is used for calculation of phonon frequencies the
results are free of anharmonic contributions by construction. Nevertheless the
results of Refs.~\cite{GGFA01p,KDJA01pv3,BHR01p} for the $E_{2g}(\Gamma)$ mode
show a large spread, too (cf.\ Tab.\ \ref{tab3}).

We note that there is still another factor which can be a source of considerable
discrepancies. The large mode Gr\" uneisen parameter implies a high sensitivity
of the calculated $E_{2g}(\Gamma)$ phonon frequency to the (equilibrium) volume
used. Some authors choose to perform the phonon calculations at the {\it
experimental} equilibrium $V_0^{exp}$ rather than at the {\it calculated}
equilibrium $V_0^{th}$. Both choices are legitimate and, as a rule, they do not
lead to any marked disagreements; the modes with large $\gamma$ may nevertheless
constitute an exception. Our present calculations refer to the {\it calculated}
equilibrium $V_0^{th}$, and the $E_{2g}(\Gamma)$ frequency at the experimental
$V_0^{exp}$ would be only 12 cm$^{-1}$ lower; the difference however becomes 35
cm$^{-1}$ in Ref.~\cite{BHR01p} where the authors were careful enough to
determine $two$ sets of frequencies for all $\Gamma$-point phonons, $viz.$ one at
the theoretical and one at the experimental equilibrium. Generally, the existence
of two results (one at $V_0^{th}$ and one at $V_0^{exp}$) is an intrinsic
uncertainty underlying the concept of frozen phonons itself \cite{rjn} and, in a
sense, it can be thought of as a consequence of the DFT-LDA. If we estimate that,
in a typical DFT-LDA calculation, the $V_0^{th}$ is between 1 and 3\% under the
experimental value $V_0^{exp}$, and with the Gr\" uneisen parameter of the order
$\gamma$ = 2.5, the two calculations of frequencies of $E_{2g}(\Gamma)$ can
differ by as much as 2.5 to 7.5\% ($\approx$ 40 cm$^{-1}$).

Another possible source of discrepancies between frozen-phonon results are the
errors of statistical nature that relate to the details of the fitting of Eq.
\ref{eqeu}, e.g., the choice of the degree of the polynomial and the selection of
the displacements $u$ for which the $E(u)$'s are calculated and the polynomial
then fitted. Even if the consequent differences are relatively small, both
choices should be clearly stated for the sake of reproducibility.

The $E_{2g}(\Gamma)$ mode is, apparently, very sensitive to all
details of the calculation. Besides the convergence with the
plane-wave cutoff, the $\vec k$-point sampling, and the technique used
for dealing with the incompletely filled orbitals (the ``metallic
sampling'' algorithm) -- which are amenable to quantitative testing --
still another source of divergences between different calculations is
in the choice of the pseudopotentials or of the method for
determination of the total energy. The consequent uncertainties are
difficult to estimate though.

Considering the large spread, among different authors, in the
calculated values of the E$_{2g}(\Gamma)$ mode frequencies,
experimentally, one might attempt to discern this mode from the other
ones by its unusually large value of the Gr\" uneisen parameter
$\gamma$ rather than by its frequency.

\subsection{Mode Gr\"{u}neisen parameters and pressure dependence of $T_c$}

The whole optical phonon branch connecting the E$_{2g}(\Gamma)$ and the
E$_{2u}(A)$ modes was shown to exhibit strong electron-phonon coupling
\cite{KDJA01pv3,BHR01p}. Within a me\-di\-um or strong coupling
scenario it is considered to largely determine the superconducting
properties of MgB$_2$. Based on early calculations of the
electron-phonon coupling constant $\lambda \approx 0.7$, an assumed
mode Gr\"{u}neisen parameter $\gamma = 1$, and a calculated pressure
dependence of the density of states at the Fermi level of $d \ln
N(0)/d P = -0.31$\%/GPa, the pressure dependence $d \ln \Tc/d P$ of
the critical temperature was estimated to fall in the range $-2.3$ to
$-3.6$\%/GPa \cite{LS01}. The estimate compared well with experimental
data and thus supported the notion of BCS superconductivity in \MB.
The pressure-induced increase in characteristic phonon frequency was
identified as the dominant cause of the decrease of $T_c$ under
pressure.

With a refined calculated value of $\lambda=0.87$ \cite{KDJA01pv3},
$\omega_0=500$~\WN, and a typical Coulomb pseudopotential of $\mu^*=0.1$ we
obtain $T_{c,\,P=0} =36.7$~K based on the McMillan expression in the form
given in Ref.~\cite{LS01}. Following the analysis of Ref.~\cite{LS01} but
using an average $\gamma = 2.3$ for the $E_{2g}(\Gamma)$--$E_{2u}(A)$ phonon
branch (taking into account anharmonicity) and a bulk modulus $B_0=145$~GPa
as calculated above, as well as $d\ln N/d P=-0.31$\%/GPa \cite{LS01}, we
estimate $d T_c/d P
\approx -1.8$~K/GPa. Experimental values of $d T_c/d P$ in the range $-0.7$
to $-2.0$~K/GPa have been reported
\cite{STII01,MNRR01,LMC01pv2,THSH01pv2,LMC01ap}. The large spread of
experimental data was, at least in part, attributed to differences in the
samples studied \cite{LMC01ap}. Our estimate of the pressure dependence of
$T_c$ is close to the upper limit of the experimental values. This, however,
is not surprising because we used the large Gr\"{u}neisen parameter of the
dominant $E_{2g}(\Gamma)$--$E_{2u}(A)$ phonon branch. Non-negligible
contributions from other vibrational modes with smaller pressure dependences
are expected to reduce the pressure shift of $T_c$ compared to our estimate.

Other effects may also be of relevance, especially a possible pressure
dependence of the electron-ion matrix element $I$ which enters the
electron-phonon coupling constant $\lambda$ and was assumed to be
constant here. Nevertheless, the present results (in particular the
large mode Gr\"{u}neisen parameter of the phonon branch with the strongest
electron-phonon coupling) underline the dominant role of the pressure
dependence of the phonon spectrum with regard to $T_c$.

\section{Experiments}
\label{sec:exp}

\subsection{Experimental Details}

Polycrystalline MgB$_2$ was prepared from stoichiometric mixtures of Mg (Johnson
Matthey Inc., 99.98\%) and natural Boron powder ($\sim$325 mesh, 99.99\%,
Aldrich). The reaction was carried out at 850$\degree$C for two days using sealed
Ta capsules under Ar atmosphere that were in turn encased in evacuated silica
tubes. After grinding under Ar, the sample was annealed at the same temperature
for one day. The x-ray powder diffraction pattern showed a single phase sample
with $a=308.63(3)$~pm and $c = 352.34(3)$~pm. In susceptibility measurements the
superconducting transition was observed at $T_c = 38.2(1)$~K.

For the high-pressure experiments at ambient temperature small amounts
of the powder samples were placed into gasketed diamond anvil cells
(DACs). The remainder of the sample cavity was filled with KCl as a
quasi-hydrostatic pressure medium. The sample was in direct contact
with the diamond window through which the Raman and reflectance spectra
were taken.

High-pressure Raman spectra were excited at 633~nm employing a
long-distance microscope objective. They were recorded in
back-scattering geometry using a single-grating spectrometer with a
multi-channel CCD detector and a holographic notch filter for
suppression of the laser line (Dilor LabRam). For the Raman
experiments the DAC was equipped with synthetic diamonds (Sumitomo type
IIa) which emit only minimal luminescence. Optical reflectivity
spectra of \MB\ in the energy range 0.6--4.0~eV were measured using a
micro-optical setup described in Ref.~\onlinecite{GS98}. Pressures
were measured by the ruby luminescence method \cite{PBBF75,MXB86}.
Low-temperature Raman spectra of \MB\ were excited with the 647-nm
line of a Kr-ion laser. The scattered light was analyzed by a
triple-grating spectrometer (Jobin-Yvon T64000) in combination with a
multi-channel CCD detector. A DAC with synthetic diamonds was used as a
low-luminescence container for the powder sample.

\subsection{Optical reflectance under pressure}

We first take a brief look at the optical reflectance spectra of MgB$_2$, because
they provide some information for understanding the evolution of the Raman
spectra under pressure. At low pressures, a low reflectance is observed
throughout the spectral range from 0.6 to 4 eV [Fig.~\ref{fig:Reflectance}(a)].
At higher pressures the NIR reflectance increases and above 10 GPa we observe a
Drude-like edge characteristic of a metallic sample. This edge then remains
present even upon pressure release from 25 to 1~GPa. This observation indicates
that the grains of the \MB\ powder may be covered with some non-metallic surface
layer which is, at least partially, removed by the deformation of the grains upon
compression. X-ray photoemission studies have indeed shown that \MB\ samples
which have been exposed to air are covered with surface layers of B$_{2}$O$_{3}$
\cite{CLWZ01p} and possibly MgCO$_{3}$ or Mg(OH)$_{2}$ \cite{VJPK01pv2}. Werheit
\etal\ concluded that samples of boron-rich compounds in general tend
to be covered with surface contaminants which are not transparent near
500~nm \cite{WSKK99}. Consequently, Raman spectra of such compounds
excited in the visible spectral region may be dominated by
contributions from the surface layer rather than the bulk material.

The energetic position of the Drude edge uncovered by pressure cycling is
probably determined by the onset of strong interband absorption at energies
around 2 eV. The fact that the edge shows a small red shift of less than
10~meV/GPa on releasing the pressure from 25 to 1~GPa indicates a small change of
the related interband absorption threshold. We have calculated the interband
contributions $\varepsilon_{xx}(\omega)$ and $\varepsilon_{zz}(\omega)$ to the
imaginary part of the dielectric function of MgB$_2$ [see Fig.
\ref{fig:Reflectance}(b)] using the WIEN97 code \cite{wien} with parameters as
described in Ref.~\cite{LS01}. Without going into details here, we just note that
the calculated $\varepsilon_{xx}(\omega)$ indeed shows a threshold at 2.1 eV,
which shifts under pressure to higher energy at a rate of 8 meV/GPa. These
interband transitions may result in resonance effects in Raman scattering.

\subsection{Raman spectra under pressure}

Figure \ref{fig:P-Spex} shows ambient-temperature Raman spectra of \MB\
recorded at increasing pressures in the range 0--14~GPa and decreasing
pressures of 15--0~GPa. In between the pressure was increased to
$\sim$20~GPa. The 0.6-GPa spectrum of Fig.~\ref{fig:P-Spex}(a), consisting
of a broad peak near 600~\WN\ with a high-energy shoulder, is representative
for the ambient-pressure Raman spectra of numerous samples we have
investigated. It is also quite similar to an ambient-pressure Raman spectrum
of \MB\ reported by Bohnen \etal\ \cite{BHR01p}. With increasing pressure
the shoulder gains intensity and the spectrum exhibits two clearly resolved
peaks. The enhanced intensity of the higher-energy peak persisted upon
pressure release until the cell was opened and the sample was exposed to air
[Fig.~\ref{fig:P-Spex}(b)].

Neither of the observed peaks can be attributed to the Raman-active
$E_{2g}(\Gamma)$ mode nor to the silent $B_{1g}(\Gamma)$ mode. The observed
pressure-induced peak shifts [Fig.~\ref{fig:P-PPos}, Tab. \ref{tab4}]
translate to mode Gr\"{u}neisen parameters of $\gamma_1 = 0.27$--0.45 and
$\gamma_2 = 0.8$--1.2 for peaks (1) and (2), respectively, using $B_0 =
145$~GPa. Hence, the calculated Gr\"{u}neisen parameter of the $E_{2g}(\Gamma)$
phonon and the experimental value for peak~(1) differ by a factor of
$\sim$6. Similarly, for peak~(2) the mode Gr\"{u}neisen parameter and the 0-GPa
frequency differ by a factor 2.3 and $>$200~\WN, respectively, from the
values calculated for the $E_{2g}(\Gamma)$ mode. The pressure-induced peak
shifts appear not to be fully reversible. This may, at least in part, be
attributed to difficulties in determining the exact peak positions because
of an intense nonlinear background that changes with pressure. It does not,
however, compromise the above conclusion.

Tentatively, we attribute the higher-energy mode~(2) with $\omega_0 =
750 \pm 20$~\WN\ to a well-defined peak in the phonon density of
states near 730~\WN\ according to calculations by Kong \etal\
\cite{KDJA01pv3} and Bohnen \etal\ \cite{BHR01p} as well as inelastic
neutron scattering data by Osborn \etal\ \cite{OGKH01p}. Observation
of this density-of-states peak in our spectra implies a violation of
the momentum conservation for the Raman scattering process which may
originate from disorder of the sample and/or the small optical
penetration depth of the laser light into the {\it metallic} sample. A
violation of the momentum conservation would not lead to spectral
features near the frequency of the $E_{2g}(\Gamma)$ mode, as there is
no density-of-states peak at this frequency\cite{KDJA01pv3,BHR01p}.

By taking into account the observations made in reflectance
measurements we conclude that only the higher-energy peak~(2) in our
Raman measurements originates from \MB\ which is uncovered from surface
layers through the application of high pressure. The lower-energy
peak~(1) which is the dominant feature of the Raman spectra of \MB\ at
ambient conditions is attributed to a surface layer of different
chemical composition. It should be noted that we observed this mode
also for samples that had been stored under inert conditions.
Probably, the surface layers form already during the material
synthesis.

Goncharov \etal\ have also studied \MB\ at high pressures by Raman spectroscopy
\cite{GSGH01pv2}. In contrast to our results they observed only a single broad
spectral feature (FWHM $\sim$300~\WN) centered near 620~\WN\ (at 0~GPa) showing a
large pressure dependence ($\gamma = 2.9$). They assign the feature to
first-order Raman scattering, but also consider other possible interpretations.
Their observations may be reconciled with ours if one assumes that the two
components visible in our spectra (FWHM $\sim$180~\WN) contribute to the very
broad spectral feature observed by Goncharov \etal\ The change in the relative
intensities of the components (1) and (2) would then lead to a large pressure
dependence of the ``smeared-out'' peak. The weighted average (``center of
gravity'') of the two peak positions exhibits an effective Gr\"{u}neisen parameter
$\gamma = 2.0$ [Fig.~\ref{fig:P-PPos}]. A direct interpolation between peak~(1)
at 0~GPa and peak~(2) at 14~GPa results in $\gamma = 4.0$. The resulting range
for the effective $\gamma$ of 2.0--4.0 thus includes the value $\gamma = 2.9$
given in Ref.~\cite{GSGH01pv2}.

\subsection{Low-temperature Raman spectra}

We investigated whether it is possible to observe a pair-breaking
excitation below $\Tc$ similar to what has been reported for
conventional superconductors such as Nb$_{3}$Sn and V$_{3}$Si
\cite{DKWF83}.

Upon cooling of a \MB\ sample below 50~K we observed a buildup of additional
scattering intensity in the range 50--300~\WN\ [Fig.~\ref{fig:TT-Spex}(a)]
which can be modeled by two Gaussian peaks located at 128 and 226~\WN,
respectively, in the 2-K spectrum. These peak positions are higher in energy
than $2\Delta_0 \approx 90$~\WN\ according to BCS theory, i.e.\ $2\Delta_0
\approx 3.5 k_B T_c$ with $T_c = 38$~K. A number of experiments, however, have
indicated superconductivity in \MB\ to be in the medium or strong-coupling regime
with $2\Delta_0 / k_B T_c$ up to 5 ($2\Delta_0 \approx 130$~\WN) \cite{KIKM01p}.
With increasing temperature these peaks vanish only at a temperature between 74
and 100~K, i.e., well above $\Tc$ [Fig.~\ref{fig:TT-Spex}(b)]. They were observed
only for parallel polarizations of the incident and scattered light. Furthermore,
the rising intensity below 30~\WN\ is not due to residual stray light but it
originates from Raman scattering by the sample. The physical origin of the
reported spectral features is unclear at present.

Chen \etal\ also investigated the superconducting gap of \MB\ by Raman
spectroscopy \cite{CKIL01p}. They reported a redistribution of
spectral weight at frequencies below $\sim$200~\WN\ and the appearance
of a small peak at 110~\WN\ which they attributed to the pair-breaking
excitation. Differing from our results a polarization dependence of
the Raman spectra was not observed.

\section{Conclusions}
\label{sec:concl}

Using the frozen phonon approach we have calculated the harmonic frequencies of
selected optical phonons of MgB$_2$ for different volumes corresponding to a
pressure range of 0 to 30 GPa. Large differences are found in the absolute values
of mode Gr\"{u}neisen parameters. In particular, for the whole
$E_{2g}(\Gamma)$--$E_{2u}(A)$ branch the frequencies are highly sensitive to
volume changes, as indicated by large values of the corresponding mode Gr\"{u}neisen
parameters. In earlier calculations\cite{KDJA01pv3,BHR01p} the
$E_{2g}(\Gamma)$--$E_{2u}(A)$ branch of the phonon dispersion was identified as
the one which exhibits the strongest electron-phonon coupling. Therefore, within
a me\-di\-um or strong coupling scenario it is considered to govern the
superconducting properties of MgB$_2$. Within this picture the present results
for the mode-Gr\"{u}neisen parameters underline the dominant role of the phonon
spectrum with respect to the observed decrease of $T_c$ with pressure.

Based on our calculations and Raman scattering experiments on \MB\ at high
pressures we conclude that the dominant Raman peak near 600~\WN\ in the
ambient-pressure spectra reported by various authors does not originate from
\MB. We rather relate it to a contaminent phase present at the sample surface. A
second peak located near 750~\WN, visible only as a weak shoulder in spectra
of as-grown samples, is tentatively attributed to a peak in the phonon
density of states. A calculation of the phonon density of states  {\it under
pressure}, extending previous ambient-pressure work \cite{KDJA01pv3,BHR01p},
is desirable to test this assignment.

Temperature-dependent Raman experiments on \MB\ showed a build-up of additional
scattering intensity in the region below 300~\WN\ upon cooling below $T_c$. A
broad double-peak structure appeared at 130/230~\WN\ and vanished only upon
warming the sample to temperatures well above $T_c$. Further study of the
low-temperature Raman spectra of \MB\ is clearly needed to elucidate the physical
origin of the various observations.

\acknowledgments

We acknowledge useful discussions with R. Zeyher. The computer
resources used in this work were provided by the Scientific Committee
of IDRIS, Orsay (France).




\begin{thebibliography}{10}
\bibitem{foot1}
Only data from the interval $V$ = (26.8, 29.6) \AA$^3$ were used for
  determination of static equilibrium; the remaining ones ($V < 26.8$) provide
  an independent checkout.
\bibitem{foot2}
Within the LDA (i.e. without the GGA), we obtained $V_0^{th}$ = 27.371 with
  2$p^6$ treated as valence electrons and $V_0^{th}$ = 27.421 without the
  2$p^6$: -5.3\% off the experimental value.
\bibitem{foot3}
The detailed exploration of $E(u)$ at very small displacements ($\vert u/a
  \vert < 0.010$) was carried out for the $E_{2g}(\Gamma)$-(b) mode only.
\end{thebibliography}


\onecolumn
\clearpage
\widetext

%
%
%
%

\begin{table}
\caption{Calculated structural parameters of MgB$_2$ at zero pressure:
equilibrium volume ($V_0$), lattice constants ($a_0, c_0$), bulk modulus ($B_0$)
and its pressure derivative ($B'$). Variation of the calculated $c/a$ ratio with
volume is given by the quadratic polynomial $c/a = c_0/a_0 + \alpha (1 - V/V_0) +
\beta (1 - V/V_0)^2 $ with the coefficients $\alpha$, $\beta$ listed below. }
\label{tab1}
\smallskip

\begin{tabular}{lcccccccc}

                           & $V_0$     & $a_0$ & $c_0$  & $B_0$  &$B'$ & $c_0/a_0$ & $\alpha$ & $\beta$ \\
                           & (\AA$^3$) & (\AA)     & (\AA)    &  (GPa)   &    &     &  & \\
   \hline Present calc.    & 28.663    & 3.065  & 3.522 & 144.7 & 3.6 &1.1490 & $-$0.258 & $-$0.06 \\
   Experiment              & 28.917(1)\tablenotemark[1] &3.08230(2)\tablenotemark[1] &
3.51461(5)\tablenotemark[1] & 147--155\tablenotemark[2] & (4.0\tablenotemark[2])
& 1.14026\tablenotemark[1] & &  \\

Difference & $-$0.9\%  & $-$0.6\% & $+$0.2\%   &   \    & \ & \  & \ &       \\
   Previous calc.~\cite{LS01} & 28.888 & 3.075  & 3.527  & 140.1  &  3.93  &
   1.1468   & $-$0.211 & $-$0.272 \\
\end{tabular}

\tablenotetext[1]{Ref.~\cite{JHS01p}, at 37 K.}
\tablenotetext[2]{Refs.~\cite{VSHY01p,JHS01p,GSGH01pv2} with the {\it assumption} $B'=4$.}
\end{table}


\widetext

\begin{table}
   \caption{Calculated pressure and volume dependence of selected phonon
   frequencies in MgB$_2$. The zero pressure frequency $\omega_0$ and the linear
   and quadratic pressure coefficients were obtained by least square fits of
   $\omega(P) = \omega_0 + a \cdot P + b \cdot P^2$ to the calculated data, and
   the mode Gr\"{u}neisen parameters $\gamma_0$ (at equilibrium volume) are derived
   from a similar quadratic expression for $\omega(V)$. $P$ has been obtained
   from $V$ through the calculated $P-V$ relation (see text).}
\label{tab2}
\smallskip

\begin{tabular}{ldddd}
   Mode  & $\omega_0$   & $a$       & $b$      &   $\gamma_0$      \\
         & (\WN)    & (\WN/GPa) & (\WN/GPa$^2$) &          \\
\hline
$B_{1g}(\Gamma)$ & 695 & 3.06506  & -0.018995 & 0.6   \\ $E_{2g}(\Gamma)$ & 535 &
8.9744   & -0.0780   & 2.5   \\ $B_{2u}(A)$      & 636 & 1.70378  & -0.01163  &
0.4   \\ $E_{2u}(A)$      & 480 & 8.9019   & -0.0809   & 2.8   \\
\end{tabular}
\end{table}

\widetext

\begin{table}
\caption{ Harmonic contributions to the phonon frequencies
at the $\Gamma$-point calculated by different ab initio methods. All
frequencies are given in cm$^{-1}$.
 } \label{tab3}
\smallskip

\begin{tabular}{ddddll}
$B_{1g}(\Gamma)$&$E_{2g}(\Gamma)$&$A_{2u}(\Gamma)$&$E_{1u}(\Gamma)$&Method&Reference\\
\hline
 690   & 470    & 390      & 320     &Frozen Phonon&   Kortus\cite{KMBA01} \\
 679   & 665    & 419      & 328     &Linear Response&   Satta\cite{GGFA01p}  \\
 692   & 585    & 401      & 335     &Linear Response&   Kong\cite{KDJA01pv3} \\
 696   & 536    & 394      & 322
                &Linear Response& Bohnen\cite{BHR01p} - I : at $V_0^{exp}$ \\
 702   & 571    & 405      & 327
                &Linear Response& Bohnen\cite{BHR01p} - II: at $V_0^{th}$ \\
 702  & 486\tablenotemark[1]& 402 & 328&Frozen Phonon&  Yildirim\cite{YGLB01p}   \\
\hline
 695   & 535  & 400\tablenotemark[2] & 333\tablenotemark[2]
                                          &Frozen Phonon& Present work  \\
\end{tabular}
\tablenotetext[1]{Frequency based on the harmonic term of the
quadratic expansion of $\Delta E_{tot}(u)$ (analogous to Eq.
\ref{eqeu}). The ``shifted'' frequency obtained by treating the
anharmonicity within the Self-Consistent Harmonic Approach or based on
the energy levels of an anharmonic oscillator amounts to 565 or 601 cm$^{-1}$,
respectively.  }
\tablenotetext[2]{Results obtained by fitting the calculated $E(u)$
with a {\it quadratic} polynomial. }
\end{table}

\widetext

\begin{table}
\caption{Observed frequencies $\omega_0$ of Raman features of MgB$_2$ and
their pressure coefficients $a=d\omega/dP$. The values for the mode Gr\"{u}neisen
parameters $\gamma_0$ are based on a bulk modulus $B_0=145$~GPa. } \label{tab4}
\smallskip

\begin{tabular}{l@{\quad}lddd}
   Mode  &  & $\omega_0$   & $a$       &    $\gamma_0$   \\
         &  & (\WN)        & (\WN/GPa) &                 \\
\hline
   Peak (1) & $P$ up          & 597(1)  &  1.85(15)   &  0.45 \\
            & $P$ down        & 609(2)  &  1.51(19)   &  0.27   \\
            & up/down average & 603(6)  &  1.68(17)   &  0.40   \\
   Peak (2) & $P$ up          & 734(3)  &  6.3(3)     &  1.2   \\
            & $P$ down        & 770(8)  &  4.4(8)     &  0.8   \\
            & up/down average & 752(18) &  5.4(10)    &  1.0   \\
\end{tabular}
\end{table}

\clearpage

\begin{figure}
   \centering
   \includegraphics[width=0.7\hsize,clip]{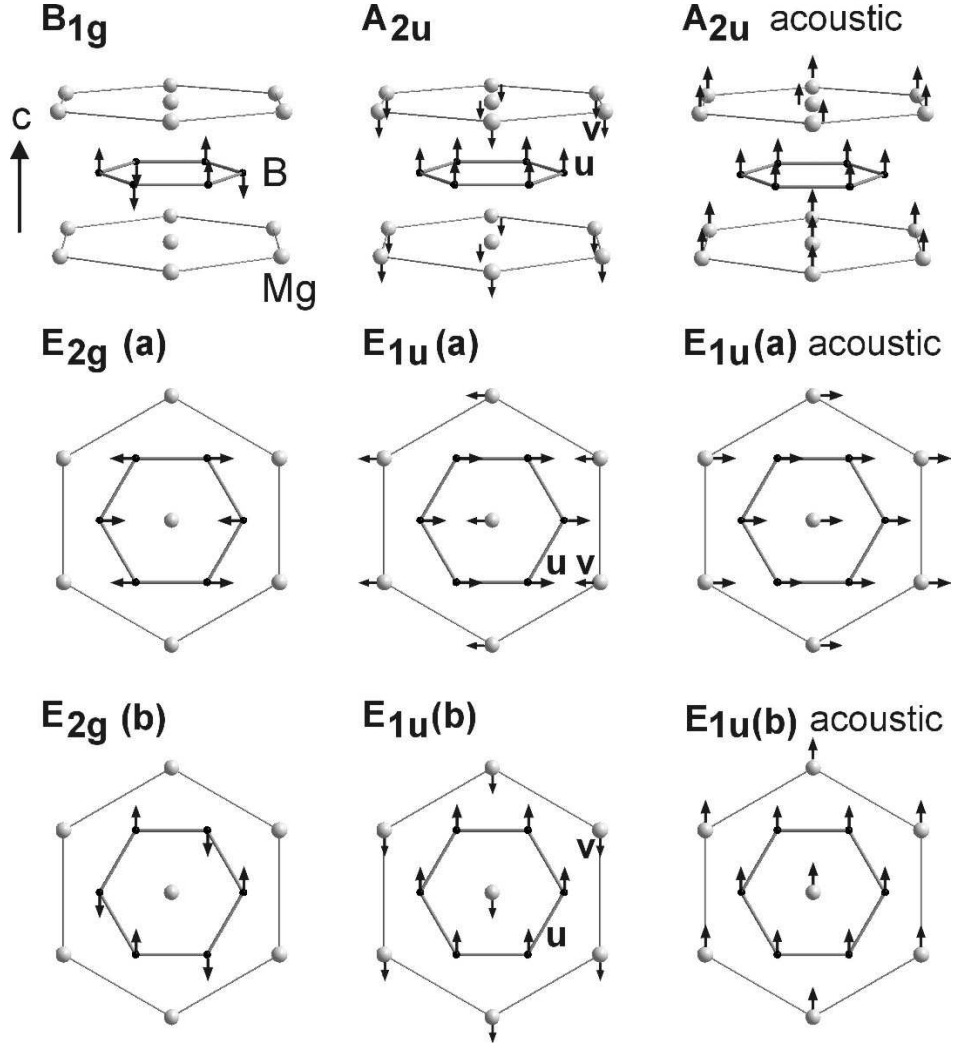}
   \vspace*{2ex}
   \caption{
         Phonon displacement patterns of the zone-center ($\Gamma$)
         modes in $MgB_2$
         (the $AlB_2$ structure, point group $D_{6h}$),
         in the order of decreasing frequencies (at zero pressure,
         columnwise).
         The labels (a), (b) denote the different patterns corresponding
         to the degenerate $E$-modes.
         For the (optic) $A_{2u}$ and $E_{1u}$ patterns to be eigenmodes, the
         displacements $u, v$ of the two sublattices
         have to be such that the unit cell's center of mass stays in rest,
         i.e. $ u : v = M_{Mg} : 2 M_B$.
         The $E_{2g}$ mode is Raman active,
         the $A_{2u}$ and $E_{1u}$ infrared active,
         the $B_{1g}$ is silent.
   }
\label{fig1}
\end{figure}

\clearpage

\begin{figure}
   \centering
   \includegraphics[width=0.7\hsize,clip]{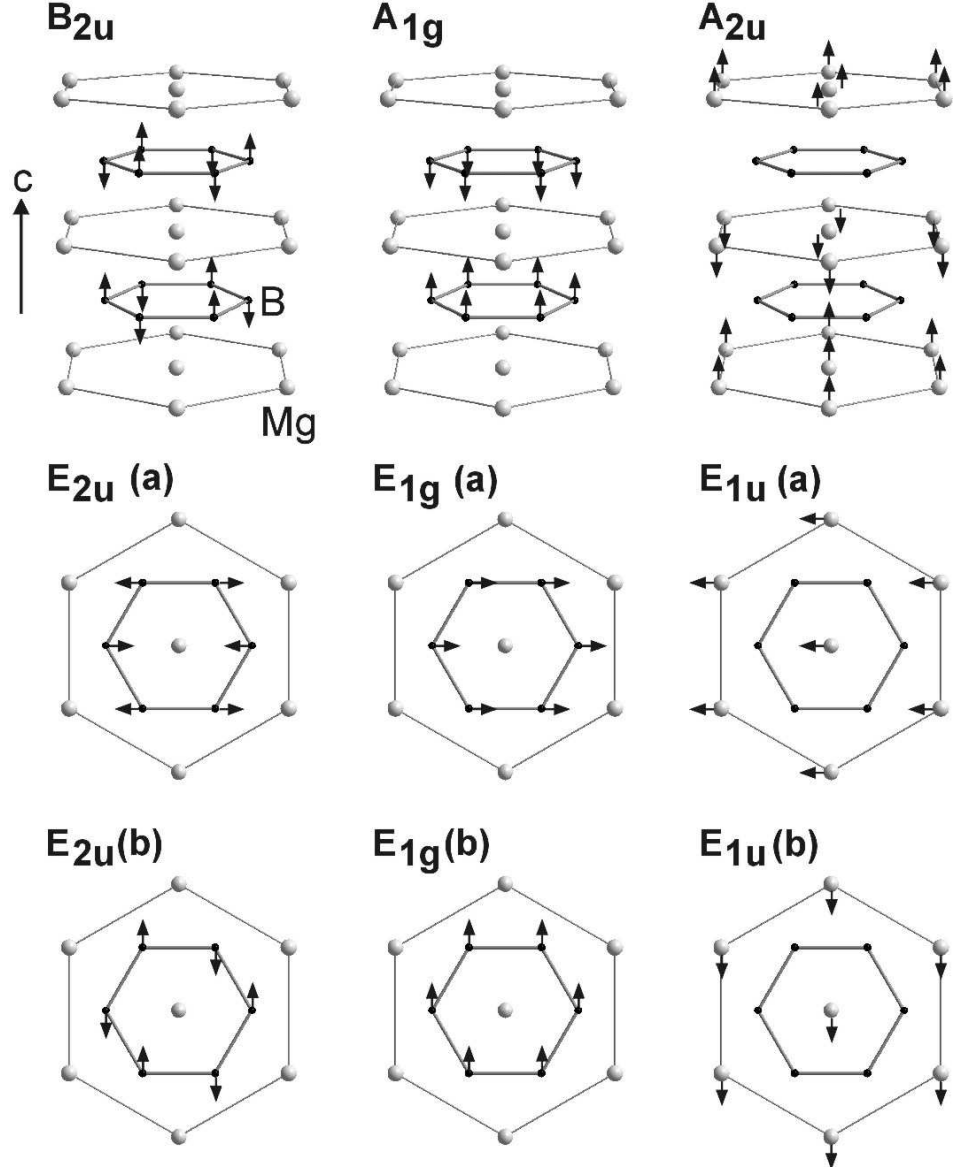}
   \vspace*{2ex}
   \caption{
         Symmetry determined displacement patterns
         for the zone-boundary modes at the $A$-point of the hexagonal
         Brillouin zone.
         Only one layer of MgB$_2$ is shown in the $E$-patterns, and
         it is understood that equivalent atoms above and below the
         layer displayed vibrate in the opposite direction.
         The patterns are arranged so as to belong to
         the same branch of the phonon dispersion
         $\omega_j(\vec k)$ as the corresponding $\Gamma$-modes in Fig. 1;
         thus e.g.\ the
         three patterns in the last column are the ``end-points''
         of the acoustic branches.}
\label{fig2}
\end{figure}

\clearpage

\begin{figure}
   \centering
   \includegraphics[width=0.7\hsize,clip]{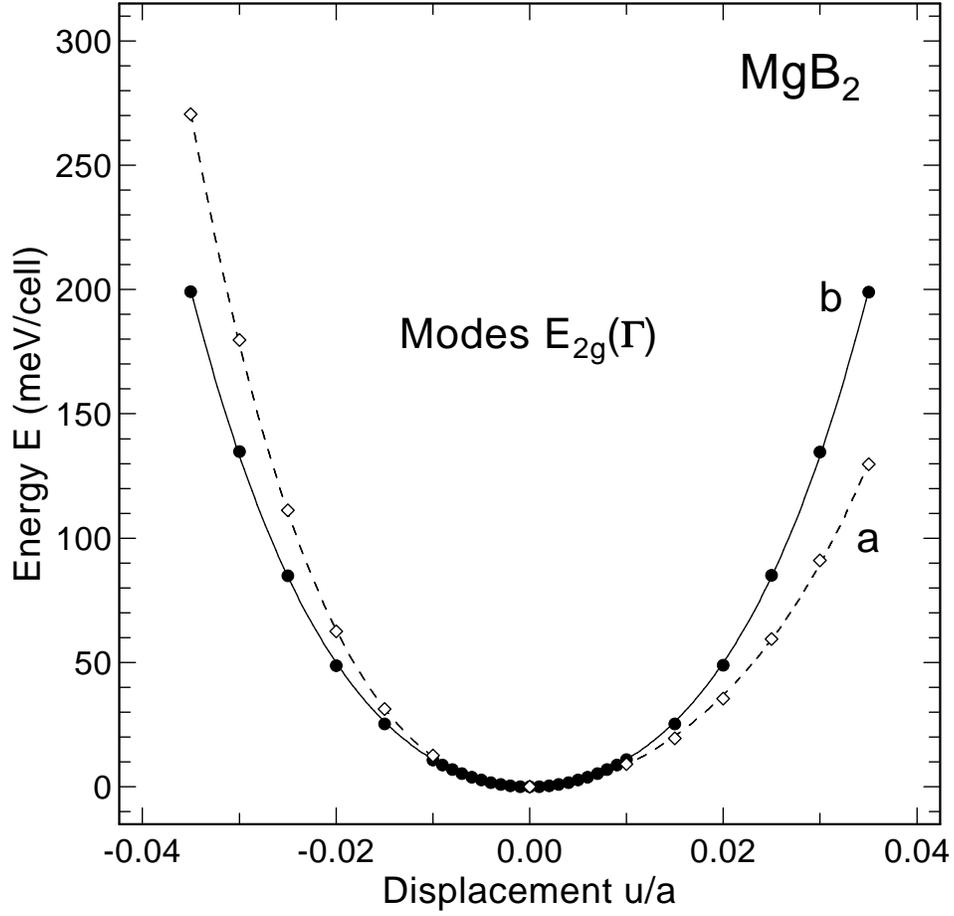}
   \vspace*{2ex}
   \caption{Calculated total energy of MgB$_2$ at equilibrium
   volume as a function of boron displacements corresponding to
   the $E_{2g}(\Gamma)$-(a) and $E_{2g}(\Gamma)$-(b) modes of Fig. 1.
   Full circles
   and open diamonds: ab initio total energy calculations; solid and dashed
   lines: quartic polynomials, Eq. \ref{eqeu}. }
   \label{fig3}
\end{figure}

\clearpage

\begin{figure}
   \centering
   \includegraphics[width=0.7\hsize,clip]{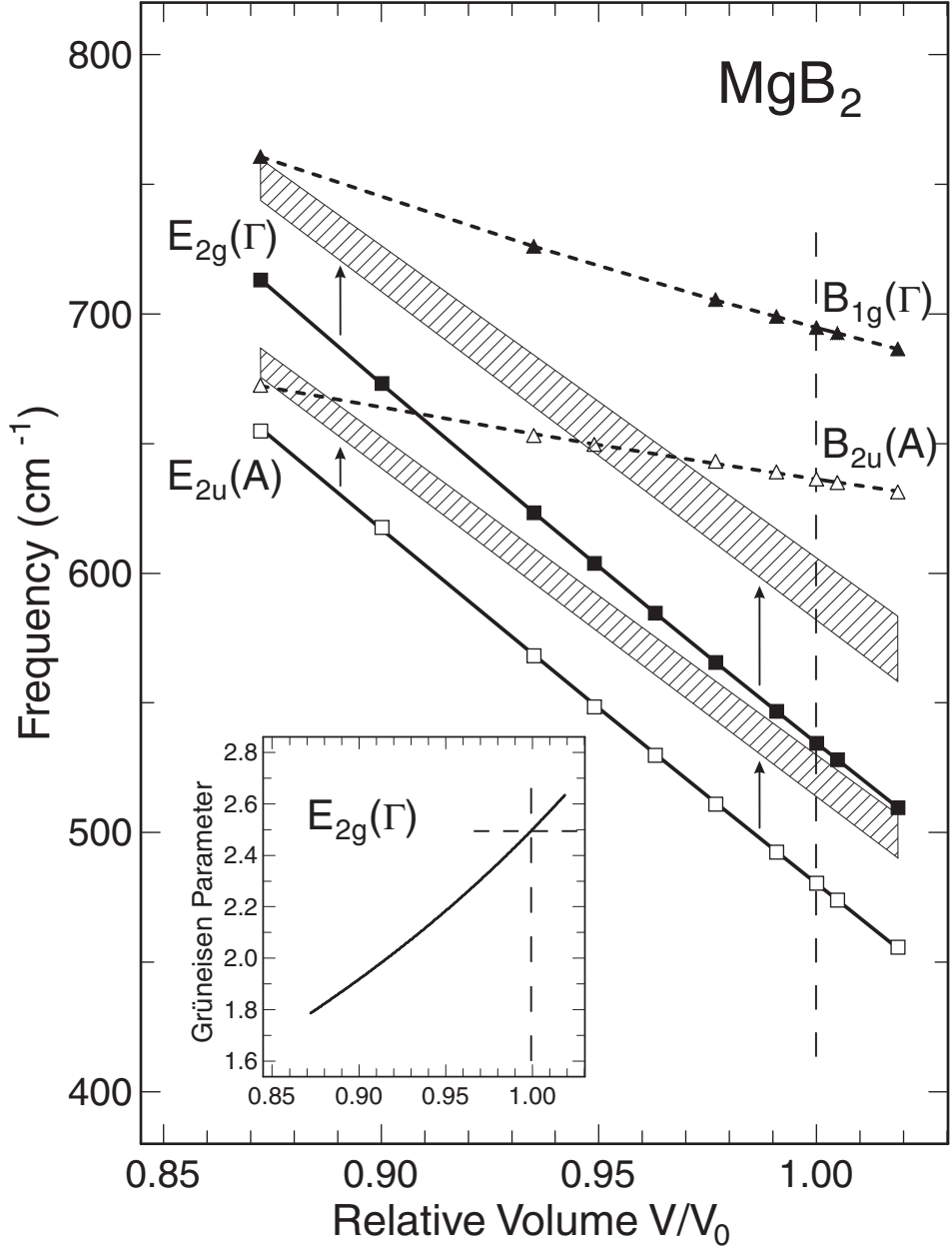}
   \vspace*{2ex}
   \caption{ Calculated harmonic frequencies of the two highest modes at
   $\Gamma$ and $A$ in  MgB$_2$ as a function of relative volume; $V_0$
   refers to calculated static equilibrium (zero pressure). The lines
   represent quadratic relation fitted to the calculated points, and they
   define the mode-Gr\"uneisen parameters $\gamma_0$ listed in Table I. Note
   that $\gamma$ itself depends on pressure, its variation with volume is
   shown in the inset. The hatched areas represent the estimated frequencies of
   the modes $E_{2g}(\Gamma)$ and $E_{2u}(A)$  when anharmonic effects
   are taken into account.}
\label{fig4}
\end{figure}

\clearpage

\begin{figure}[bt]
     \centering
     \includegraphics[width=\hsize,clip]{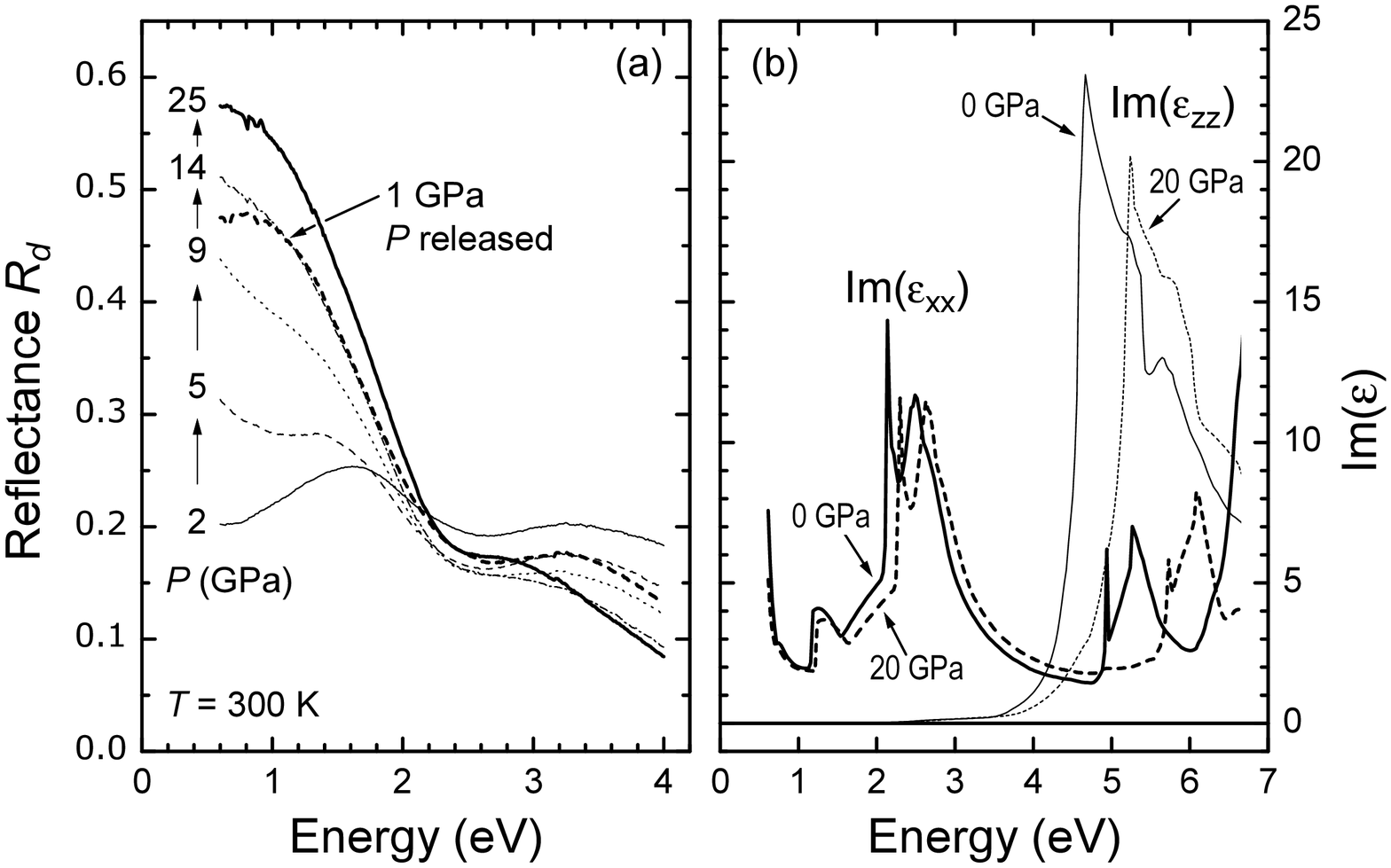}
     \vspace{1mm}
      \caption{(a) Optical reflectance spectra of \MB\ at ambient
      temperature and pressures up to 25~GPa. $R_d$ denotes the absolute
      reflectance at the diamond--sample surface. (b) Calculated imaginary
      part of the dielectric function (interband transitions) at 0~GPa
      (solid lines) and 20~GPa (broken lines) for light polarized
      perpendicular ($\varepsilon_{xx}$) and parallel ($\varepsilon_{zz}$)
      to the $c$ axis.}
     \label{fig:Reflectance}
\end{figure}

\clearpage

\begin{figure}[bt]
     \centering
     \includegraphics[width=\hsize,clip]{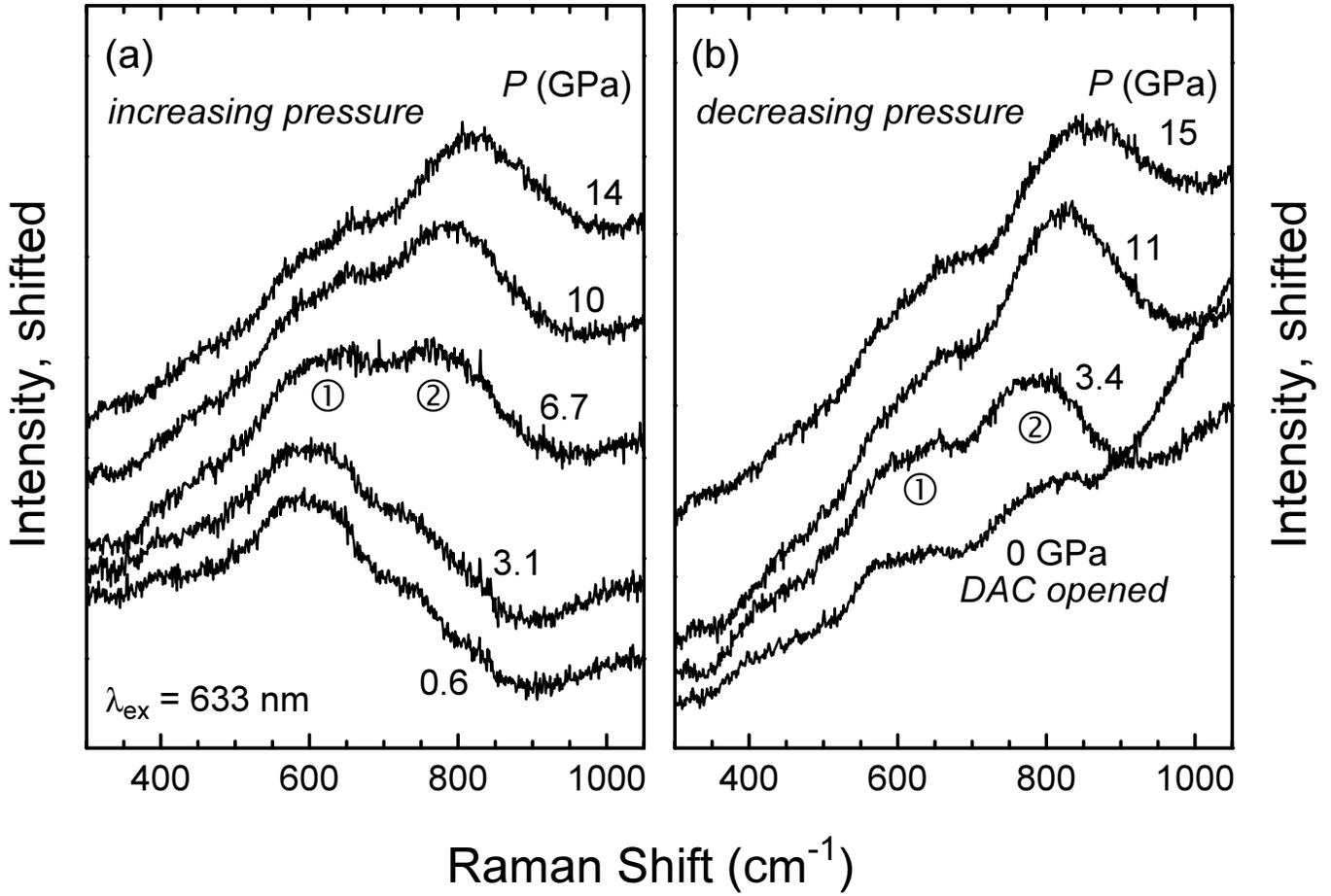}
     \vspace{1mm}
     \caption{Raman spectra of \MB\ for (a) increasing and (b) decreasing
     pressures ($T=300$~K). In between the pressure was increased to
     $\sim$20~GPa. After pressure release the sample was in contact with air
     before recording the 0-GPa spectrum in (b).}
     \label{fig:P-Spex}
\end{figure}

\clearpage

\begin{figure}[bt]
     \centering
     \includegraphics[width=0.6\hsize,clip]{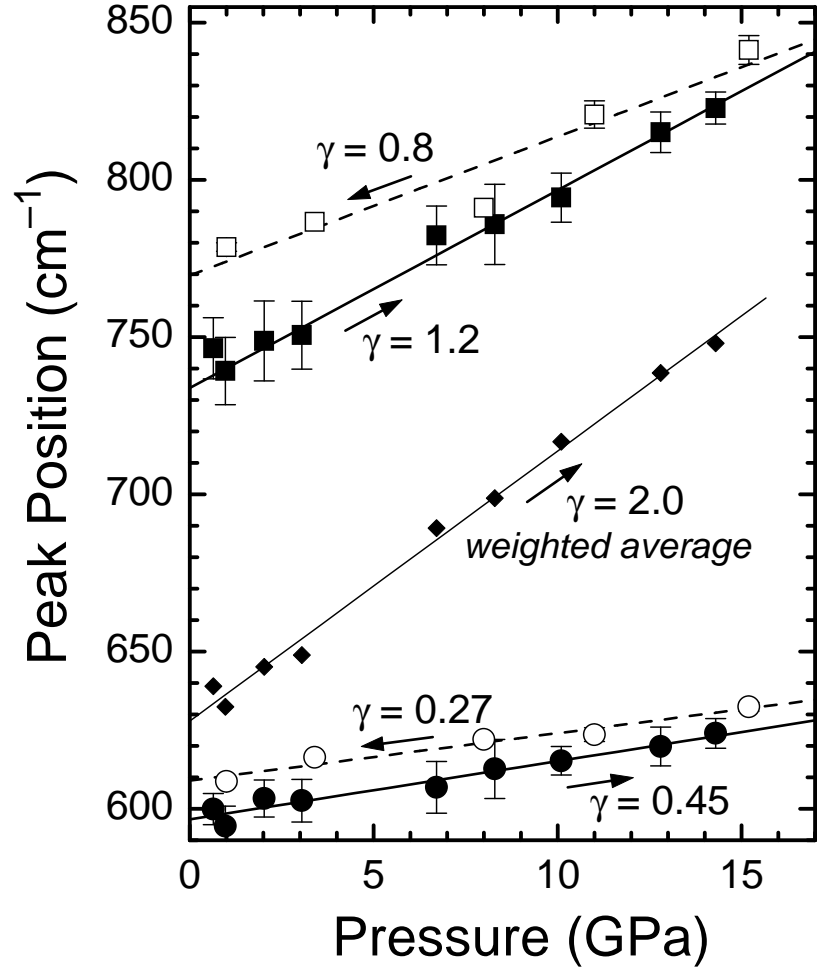}
     \vspace{1mm}
     \caption{Energies of the Raman features (1) and (2) of \MB\ as a function of
     pressure. Solid and open symbols refer to data measured at increasing
     and decreasing pressures, respectively. Lines represent linear
     relations fitted to the data.}
     \label{fig:P-PPos}
\end{figure}

\clearpage

\begin{figure}[bt]
     \centering
     \includegraphics[width=\hsize,clip]{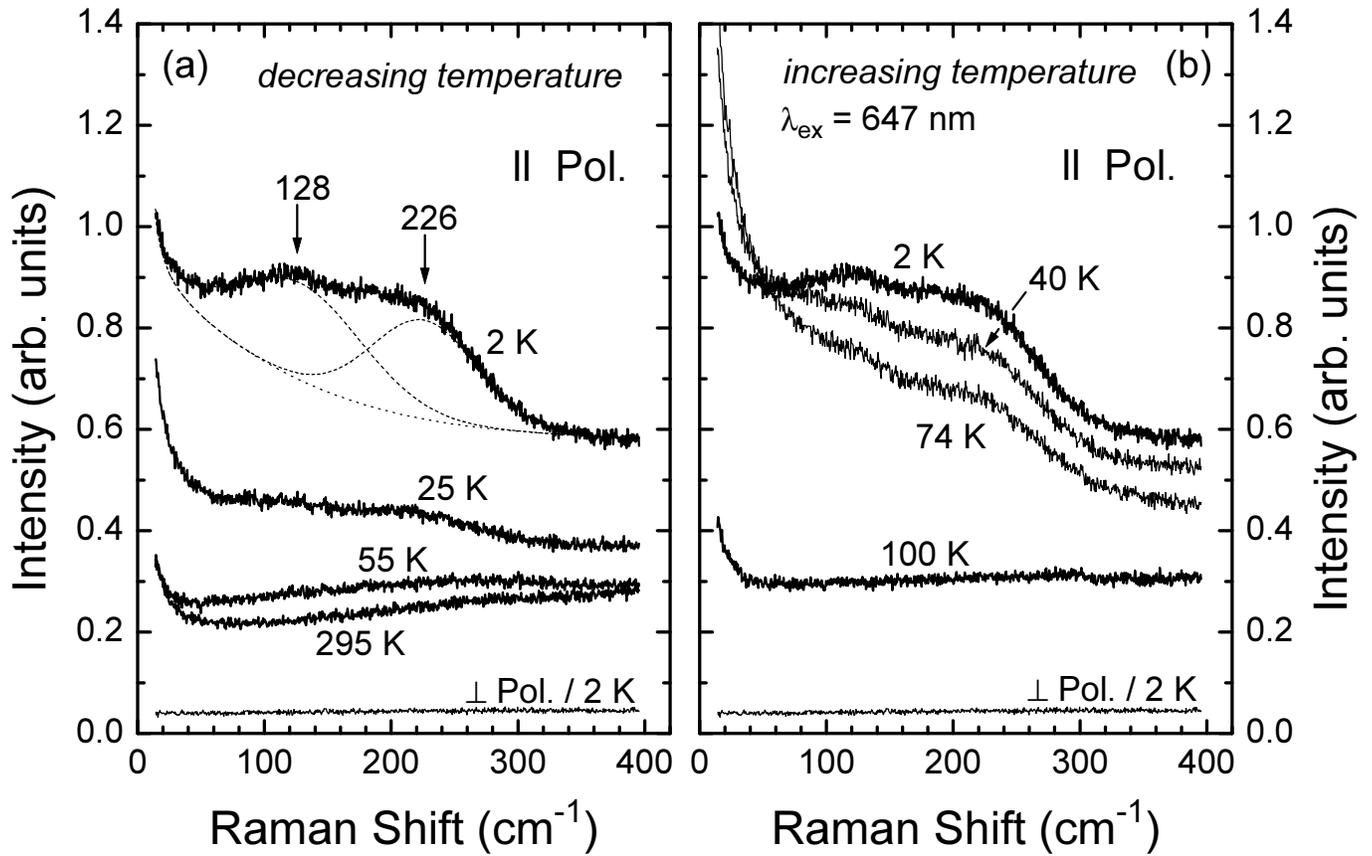}
     \vspace{1mm}
     \caption{Raman spectra of \MB\ at ambient pressure for (a)
     decreasing and (b) increasing temperatures. ``$\|$ Pol.'' and ``$\perp$
     Pol.'' refer to parallel and crossed polarizations of the incident and
     scattered light, respectively.}
     \label{fig:TT-Spex}
\end{figure}

\end{document}